\documentclass[aps,prl,showpacs,showkeys]{revtex4}
\usepackage{amsmath}
\usepackage{graphicx}
\usepackage{bm}
\usepackage{epsfig}
\begin{document}
\title{Solution of spin-boson systems in one and two-dimensional
geometry via the asymptotic iteration method}
\date{\today}
\author{Ramazan Ko\c{c}}
\email{koc@gantep.edu.tr} \affiliation{Department of Engineering
Physics, Faculty of Engineering, University of Gaziantep, 27310
Gaziantep, Turkey}
\author{Okan \"{O}zer}
\email{ozer@gantep.edu.tr} \affiliation{Department of Engineering
Physics, Faculty of Engineering, University of Gaziantep, 27310
Gaziantep, Turkey}
\author{Hayriye T\"{u}t\"{u}nc\"{u}ler}
\email{tutunculer@gantep.edu.tr} \affiliation{Department of
Engineering Physics, Faculty of Engineering, University of
Gaziantep, 27310 Gaziantep, Turkey}
\author{R. G\"{u}ler Y\i ld\i r\i m}
\email{ozkan@gantep.edu.tr} \affiliation{Department of Engineering
Physics, Faculty of Engineering, University of Gaziantep, 27310
Gaziantep, Turkey}

\begin{abstract}
We consider solutions of the $2\times 2$ matrix Hamiltonian of
physical systems within the context of the asymptotic iteration
method. Our technique is based on transformation of the associated
Hamiltonian in the form of the first order coupled differential
equations. We construct a general matrix Hamiltonian which
includes a wide class of physical models. The systematic study
presented here reproduces a number of earlier results in a natural
way as well as leading to new findings. Possible generalizations
of the method are also suggested.
\end{abstract}



\maketitle

\section{Introduction}

During the last decade, a great deal of attention has been paid to
examining different quantum optical models
\cite{prinz,wolf,winkler,wang,band}. Recently an iteration
technique \cite{ciftci1,ciftci3} has been suggested to solve the
Schr\"{o}dinger equation which improves both analytical and
numerical solutions of the problems and has been developed for
some quantum optical systems. The solution of quantum optical
Hamiltonians are important in the existing literature. In general,
the study of two level-systems in one and two-dimensional geometry
coupled to bosonic modes has been the subject of intense attention
because of its extensive applicability in various fields of
physics
\cite{judd,reik,ito,mosh,koc2,tur,jaynes,dresselhaus,rashba,tsit,frus}.
Due to the practical and technological importance of these models,
it is not surprising that various aspects have been studied both
analytically and numerically
\cite{klim,kara1,kara2,kumar,wunsce,buzek,ban,chen,balan}. Such
systems have often been analyzed using numerical methods, because
the implementation of the analytical techniques does not yield
simple analytical expressions. Remarkably, exact solutions have
not been thus far presented except for special cases, even though
it has been suggested that the problem may be solved exactly. The required
Analytical treatments need tedious calculations.

Quantum optical models are one of the most fascinating phenomena
in modern physics and chemistry, providing a general approach to
understanding the properties of molecules, crystals and their
origins. Most of these Hamiltonians have yet to be solved exactly.
Therefore, the natural question arising at this point is: can the
asymptotic iteration method be applied and used to obtain the
solutions to these systems? The answer of this question is the
main topic of this paper.

In recent years much attention has been focused on the asymptotic
iteration method (AIM)
\cite{ciftci1,sous,ciftci2,barakat1,amore,bayrak}. This method
reproduces exact solutions to many exactly solvable differential
equations and these equations can be related to the
Schr\"{o}dinger equation. It also gives accurate results for
non-solvable Schr\"{o}dinger equations, such as the sextic
oscillator, cubic oscillator, deformed Coulomb potential, etc.
which are important in applications to many problems in physics.
Encouraged by its satisfactory performance through comparison with
other methods, we feel tempted to extend AIM to solve matrix
differential equations. Although AIM has been applied to solve the
Schr\"{o}dinger equation, its application to the solution of the
matrix equations \cite{ciftci3} still needs to be improved. In
contrast to the solution of the Schr\"{o}dinger equation including
potentials of Coulomb, Morse, harmonic oscillator etc. by using
AIM, study of the matrix Hamiltonians has not attracted much
attention in the literature. Therefore, we concentrate on the
solution of a general two-dimensional two mode bosonic Hamiltonian
by using AIM in this paper.

The paper is organized as follows. In section 2 we discuss transformation of
the Hamiltonians whose original forms are given as boson and fermion
operators. We show that the Hamiltonian can be expressed as two coupled
first order differential equations in the Bargmann-Fock space. In section 3,
we develop AIM to obtain eigenvalues and eigenfunctions of the different
matrix Hamiltonians widely used in physics. Section 4 is devoted to solve a
wide class of physical Hamiltonians in the framework of the AIM. Finally we
conclude our results in section 5.

\section{Two dimensional two mode bosonic Hamiltonian}

The most general form of the Hamiltonian consists of the coupling
of a single spin-1/2 to the boson field in two dimensional
geometry. It can be written as
\begin{equation}
H=H_{0}+\omega _{0}\sigma _{0}+\left( \kappa _{1}a+\kappa _{2}a^{+}+\kappa
_{3}b+\kappa _{4}b^{+}\right) \sigma _{+}+\left( \gamma _{1}a+\gamma
_{2}a^{+}+\gamma _{3}b+\gamma _{4}b^{+}\right) \sigma _{-}  \label{e1}
\end{equation}%
where $H_{0}=\hbar \omega _{1}\left( a^{+}a\right) +\hbar \omega _{2}\left(
b^{+}b\right) $ and $\omega _{i},\kappa _{i}$ and $\gamma _{i}$ are physical
constants. The Pauli matrices $\sigma _{0,\pm }$ are given by
\begin{equation}
\sigma _{0}=\left(
\begin{array}{cc}
-1 & 0 \\
0 & 1%
\end{array}%
\right) ;\quad \sigma _{+}=\left(
\begin{array}{cc}
0 & 1 \\
0 & 0%
\end{array}%
\right) ;\quad \sigma _{-}=\left(
\begin{array}{cc}
0 & 0 \\
1 & 0%
\end{array}%
\right) ,  \label{e2}
\end{equation}%
and bosonic annihilation $a,b$ and creation $a^{+},b^{+}$ operators satisfy
the usual commutation relations%
\begin{equation}
\left[ a,a^{+}\right] =\left[ b,b^{+}\right] =1;\quad \left[ a,b\right] =%
\left[ a^{+},b^{+}\right] =\left[ a^{+},b\right] =\left[ a,b^{+}\right] =0.
\label{e3}
\end{equation}%
It is seen that the Hamiltonian (\ref{e1}) includes various
physical Hamiltonians depending on the choice of the parameters,
and it is Hermitian if $\kappa _{1}=\gamma _{2}$, $\kappa
_{2}=\gamma _{1}$, $\kappa _{3}=\gamma _{4}$ and $\kappa
_{4}=\gamma _{3}$ for real parameters. In general $H$ describes
spin phonon (photon) relaxation in the presence of a magnetic
field. It can also be used to study interaction of a two-level
atom with an electromagnetic field. As explicit examples, when
$\omega _{1}=\omega _{2}=\omega $, $\kappa _{2}=\kappa _{3}=\gamma
_{1}=\gamma _{4}=0$ and $\kappa _{1}=\kappa _{4}=\gamma
_{2}=\gamma _{3}=\kappa $, the Hamiltonian $H$ reduces to the
$E\otimes \varepsilon $ Jahn-Teller (JT) Hamiltonian
\cite{reik,koc3} and when $\omega _{1}=\omega +\omega _{c}$,
$\omega _{2}=\omega -\omega _{c}$, $\kappa _{2}=\kappa _{3}=\gamma
_{1}=\gamma _{4}=0$ and $\kappa _{1}=-\kappa _{4}=\gamma
_{2}=-\gamma _{3}=\kappa $ then the Hamiltonian $H$ becomes the
Hamiltonian of quantum dots including spin-orbit coupling \cite{rashba,koc2}%
. One can also obtain Jaynes-Cummings (JC) Hamiltonian \cite{jaynes} and
modified JC Hamiltonian \cite{koc4} as well as many other interesting
physical Hamiltonians by appropriate choices of the parameters $\omega _{i}$%
, $\kappa _{i}$ and $\gamma _{i}$ in (\ref{e1}). There exists a relatively
large number of different approaches for the solution of the eigenvalue
problem for $H\psi =E\psi $ in the literature. However, we present here a
systematic treatment for the determination of the eigenvalues and
eigenfunctions of (\ref{e1}) in the context of AIM.

We note that the Hamiltonian of a physical system is given in the
form of a differential equation in some cases. One way to obtain
the bosonic form of a Hamiltonian is to construct a suitable
differential form of the bosons. Therefore, it is worth discussing
some useful differential forms of the bosons and the connection
between them before we begin to present a procedure to solve
(\ref{e1}).

\subsection{Forms of bosons}

We are interested in the two-level systems in a one and
two-dimensional geometry whose Hamiltonians are given in terms of
bosons-fermions or matrix-differential equations. By the use of
the differential form of the operators one can easily find the
interrelation between boson-fermion and matrix differential
equation formalisms of the Hamiltonians. There are various
differential forms of the boson operators. At this point, let us
start by introducing the following differential forms of the boson
operators :
\begin{eqnarray}
a^{+} &=&\frac{\ell }{2}(x+iy)-\frac{1}{2\ell }(\partial _{x}+i\partial
_{y}),  \notag \\
a &=&\frac{\ell }{2}(x-iy)+\frac{1}{2\ell }(\partial _{x}-i\partial _{y}),
\notag \\
b^{+} &=&\frac{\ell }{2}(x-iy)-\frac{1}{2\ell }(\partial _{x}-i\partial
_{y}),  \label{e4} \\
b &=&\frac{\ell }{2}(x+iy)+\frac{1}{2\ell }(\partial _{x}+i\partial _{y})
\notag
\end{eqnarray}%
where $\ell =\sqrt{\frac{m\omega }{\hbar }}$ is the length
parameter. In principle, if a Hamiltonian is expressed by boson
operators, one could rely directly on the known formulae of the
action of boson operators on a state with a defined number of
particles without solving differential equations. Apart from the
mentioned method, the Hamiltonians can often not be solved
exactly, then we need to develop alternative methods. It is
amazingly interesting that the Hamiltonian (\ref{e1}) can be
solved within the framework of the AIM when it is transformed into
the form of the first order coupled differential equations under
the constraints that will be given in
Eqs.(\ref{e11a})-(\ref{e11d}). Now, we briefly discuss three well
known differential forms of the bosons and the relation between
them. In order to obtain different forms of the bosons we present
a transformation procedure. For the sake of simplicity we take
$\hbar =m=\omega =1$ and then $\ell$ becomes unity.

Consider the following exponential operator
\begin{equation}
\Lambda =\exp \left[ \beta \left( a^{+}b+b^{+}a\right) \right] .  \label{e5}
\end{equation}
The operator acts on the bosons as follows:
\begin{eqnarray}
\Lambda a\Lambda ^{-1} &=&a\cos \beta -b\sin \beta ;\quad \Lambda
a^{+}\Lambda ^{-1}=a^{+}\cos \beta -b^{+}\sin \beta  \label{e6} \\
\Lambda b\Lambda ^{-1} &=&b\cos \beta +a\sin \beta ;\quad \Lambda
b^{+}\Lambda ^{-1}=b^{+}\cos \beta +a^{+}\sin \beta  \notag
\end{eqnarray}%
If we set $\beta =-\pi /4$ and change the variable $y\rightarrow
iy$ in the transformation given in (\ref{e6}), we obtain the
following relations
\begin{eqnarray}
\Lambda a^{+}\Lambda ^{-1} &=&\frac{1}{\sqrt{2}}(x-\partial _{x});\quad
\Lambda a\Lambda ^{-1}=\frac{1}{\sqrt{2}}(x+\partial _{x}),  \notag \\
\Lambda b^{+}\Lambda ^{-1} &=&\frac{1}{\sqrt{2}}(y-\partial _{y});\quad
\Lambda b\Lambda ^{-1}=\frac{1}{\sqrt{2}}(y-\partial _{y}).  \label{e7}
\end{eqnarray}%
The other important differential forms of the bosons can be
obtained by transforming the bosons with the following operator:
\begin{equation}
\Gamma =\exp \left[ \frac{\alpha }{2}\left( a^{2}+a^{+2}+b^{2}+b^{+2}\right) %
\right] .  \label{e8}
\end{equation}
where $\alpha$ is the rotation angle of bosons. The action of the operator
on the bosons is given by
\begin{eqnarray}
\Gamma a\Gamma ^{-1} &=&a\cos \alpha -a^{+}\sin \alpha ;\quad \Gamma
a^{+}\Gamma ^{-1}=a^{+}\cos \alpha +a\sin \alpha .  \notag \\
\Gamma b\Gamma ^{-1} &=&b\cos \alpha -b^{+}\sin \alpha ;\quad \Gamma
b^{+}\Gamma ^{-1}=b^{+}\cos \alpha +b\sin \alpha .\quad  \label{e9}
\end{eqnarray}%
For $\alpha =\pi /4$, the boson operators take the form
\begin{eqnarray}
a &\rightarrow &\frac{1}{\sqrt{2}}\left( a-a^{+}\right) =\frac{d}{dx};\quad
a^{+}\rightarrow \frac{1}{\sqrt{2}}\left( a+a^{+}\right) =x  \notag \\
b &\rightarrow &\frac{1}{\sqrt{2}}\left( b-b^{+}\right) =\frac{d}{dy};\quad
b^{+}\rightarrow \frac{1}{\sqrt{2}}\left( b+b^{+}\right) =y  \label{e10}
\end{eqnarray}%
The last formulation is known as the Bargmann-Fock space
description of bosons \cite{bargmann} and this form plays a key
role when constructing the one-variable first order matrix
differential equation form of (\ref{e1}). When we insert the form
(\ref{e10}) in (\ref{e1}), the Hamiltonian $H$ becomes a first
order and two variable matrix differential equation. In order to
separate the variables, we look for the conserved quantity of the
system. After some treatment, we obtain the following conserved
quantities:
\begin{subequations}
\begin{eqnarray}
K_{1} &=&a^{+}a-b^{+}b-\frac{1}{2}\sigma _{0},\quad \mathrm{when}\quad
\gamma _{2}=\gamma _{3}=\kappa _{1}=\kappa _{4}=0  \label{e11a} \\
N_{1} &=&a^{+}a+b^{+}b+\frac{1}{2}\sigma _{0},\quad \mathrm{when}\quad
\gamma _{1}=\gamma _{3}=\kappa _{2}=\kappa _{4}=0  \label{e11b} \\
K_{2} &=&a^{+}a-b^{+}b+\frac{1}{2}\sigma _{0},\quad \mathrm{when}\quad
\gamma _{1}=\gamma _{4}=\kappa _{2}=\kappa _{3}=0  \label{e11c} \\
N_{2} &=&a^{+}a+b^{+}b-\frac{1}{2}\sigma _{0},\quad \mathrm{when}\quad
\gamma _{2}=\gamma _{4}=\kappa _{1}=\kappa _{3}=0  \label{e11d}
\end{eqnarray}
We note here that although we obtain four conserved quantities, $K_{1}$
conjugates with $K_{2}$ and $N_{1}$ conjugates with $N_{2}$. They can be
provided by a similarity transformation
\end{subequations}
\begin{equation}
K=K_{2}=UK_{1}U^{-1}\quad \mathrm{and}\quad N=N_{2}=UN_{1}U^{-1}
\label{ex11}
\end{equation}
where $U=\sigma _{+}+\sigma _{-}$. It is well known that if two
quantum mechanical operators commute then they have common
eigenfunctions. Thus one can write
\begin{equation}
K\left\vert n_{1,}n_{2}\right\rangle =\left( k+\frac{1}{2}\right) \left\vert
n_{1,}n_{2}\right\rangle ,\quad N\left\vert n_{1,}n_{2}\right\rangle =\left(
k+\frac{1}{2}\right) \left\vert n_{1,}n_{2}\right\rangle ,  \label{e12}
\end{equation}
and the eigenvalue problem can easily be solved in the Bargmann-Fock space.
Thus, we obtain the following expressions for the eigenfunctions of $K$ and $%
N$:
\begin{subequations}
\begin{eqnarray}
\psi (x,y) &=&x^{k}\phi \left( xy\right) \left\vert \uparrow \right\rangle
+x^{k+1}\phi \left( xy\right) \left\vert \downarrow \right\rangle ,\quad
\mathrm{for}\quad K  \label{e13c} \\
\psi (x,y) &=&x^{k+1}\phi \left( y/x\right) \left\vert \uparrow
\right\rangle +x^{k}\phi \left( y/x\right) \left\vert \downarrow
\right\rangle \quad \mathrm{for}\quad N  \label{e13d}
\end{eqnarray}
where $\left\vert \uparrow \right\rangle $ stands for the up state and $%
\left\vert \downarrow \right\rangle $ stands for the down state.
In this case one can normalize the wavefunction $\psi (x,y)$ by
using the relation
\end{subequations}
\begin{eqnarray}
\int_{\infty } e^{-W}\psi(x,y) dx dy  \notag \label{weightfunc}
\end{eqnarray}
where $W$ is the weight function and the integral is taken over
all space. The eigenfunction of the Hamiltonian can be obtained
from the relation
\begin{equation}
\left\vert n_{1},n_{2}\right\rangle =\Gamma ^{-1}\Lambda ^{-1}\psi (x,y).
\label{e14}
\end{equation}
Meanwhile we note that the conserved quantity $N$ crucially depends on the
conservation of the number of particles. The classical motion of the
particle takes place in the space of angular momentum on a sphere. However,
to obtain a physical meaning of $K$ from another perspective; when
the motion of the particle takes place on an ellipsoid, the conserved
quantity is given in the form of (\ref{e12}). We have obtained four
different conserved quantities for the Hamiltonian (\ref{e1}) depending on
the choice of parameters. However, it is fact that $K_{1}$ and $K_{2}$
can easily be mapped into each other. It implies that Hamiltonians obtained
under the constraints (\ref{e11a}) and (\ref{e11c}) correspond to physically
similar systems. We also say that $N_{1}$ and $N_{2}$ can also be
mapped into each other and the Hamiltonians obtained under constraints
(\ref{e11b}) and (\ref{e11d}) also correspond to physically similar systems.
Therefore, it is valid to discuss the solution of the Hamiltonian
(\ref{e1}) under the conditions given in (\ref{e11c}) and (\ref{e11d}).

Since Eqs. (\ref{e11a}) and (\ref{e1}) commute, then they have the same
eigenfunctions under the constraint given in (\ref{e11a}). Thus, insertion
of (\ref{e13c}) into the Hamiltonian (\ref{e1}) and using the form
(\ref{e10}), we obtain the following set of one variable coupled differential
equations:
\begin{subequations}
\begin{equation}
\left[ \left( \omega _{1}+\omega _{2}\right) z\frac{d}{dz}+\left( k+\frac{1}{%
2}\right) \omega _{1}+\omega _{0}-E\right] \phi _{1}\left( z\right) +\left[
\left( k+1+z\frac{d}{dz}\right) \kappa _{1}+\kappa _{4}z\right] \phi
_{2}\left( z\right) = 0  \label{e15a}
\end{equation}

\begin{equation}
\left[ \left( \omega _{1}+\omega _{2}\right) z\frac{d}{dz}+\left( k+\frac{3}{%
2}\right) \omega _{1}+\omega _{2}-\omega _{0}-E\right] \phi _{2}\left(
z\right) +\left[ \gamma _{3}\frac{d}{dz}+\gamma _{2}\right] \phi _{1}\left(
z\right)  = 0  \label{e15b}
\end{equation}
where $z=xy$ and $E$ is the eigenvalue of the Hamiltonian $H$ and $\phi
_{1}\left( z\right)$ and $\phi _{2}\left( z\right)$ correspond to up and down
eigenfunctions of the Hamiltonian $H$, respectively. Similarly, when we
substitute (\ref{e13d}) into the Hamiltonian (\ref{e1}) with the form
(\ref{e10}), we obtain the following set of one variable coupled
differential equations:
\end{subequations}
\begin{subequations}
\begin{equation}
\left[ \left( \omega _{2}-\omega _{1}\right) z\frac{d}{dz}+(k+1)\omega
_{1}+\omega _{0}-E\right] \phi _{1}\left( z\right) +\left[ \kappa
_{2}+\kappa _{4}z\right] \phi _{2}\left( z\right) =0  \label{e16a}
\end{equation}

\begin{equation}
\left[ \left( \omega _{2}-\omega _{1}\right) z\frac{d}{dz}+k\omega
_{1}-\omega _{0}-E\right] \phi _{2}\left( z\right) +\left[ \gamma
_{1}(k+1)+\left( \gamma _{3}-\gamma _{1}z\right) \frac{d}{dz}\right] \phi
_{1}\left( z\right) =0.  \label{e16b}
\end{equation}
where $z=y/x$. Our task is now to apply the AIM to solve the corresponding
Hamiltonian.

\section{Development of the AIM for matrix Hamiltonians}

\label{secaim}

In this section we systematically present a procedure for the solution of $%
2\times \ 2$ first-order matrix differential equations. Consider the
following first order matrix differential equation:
\end{subequations}
\begin{equation}
\phi ^{\prime }=u_{0}\phi  \label{e19}
\end{equation}%
where $\phi =\left[ \phi _{1},\phi _{2}\right] ^{T}$, two component column
vector $u_{0}$ is a $2\times \ 2$ matrix function. Note that $\phi$ and $u_{0}$
are functions of $z$ and $\phi^{\prime}$ is the first derivative with
respect to $z$. Now, in order to obtain a general solution to this equation
in the framework of the AIM we use similar arguments to those given in \cite
{ciftci3}. The differential equation (\ref{e19}) can be written as the two
coupled equation
\begin{equation}
\phi _{1}^{\prime }=a_{0}\phi _{1}+b_{0}\phi _{2};\quad \phi _{2}^{\prime
}=c_{0}\phi _{2}+d_{0}\phi _{1}  \label{e20}
\end{equation}
where $a_{0}, b_{0}, c_{0}$ and $d_{0}$ are elements of the matrix $u_{0}$.
It is easy to show that the $n^{th}$ derivative of $\phi_{1}$ and $\phi _{2}$
can be written as
\begin{eqnarray}
\phi _{1}^{\prime \prime } &=&a_{1}\phi _{1}+b_{1}\phi _{2} ;\quad \phi
_{2}^{\prime \prime }=c_{1}\phi _{2}+d_{1}\phi _{1}  \notag \\
\phi _{1}^{\prime \prime \prime } &=&a_{2}\phi_{1}+b_{2}\phi_{2} ;\quad \phi
_{2}^{\prime \prime \prime}=c_{2}\phi_{2}+d_{2}\phi_{1}  \notag \\
&&\cdots  \label{e21} \\
\phi _{1}^{(n)} &=&a_{n-1}\phi _{1}+b_{n-1}\phi _{2} ;\quad
\phi_{2}^{(n)}=c_{n-1}\phi _{2}+d_{n-1}\phi _{1}  \notag \\
\phi _{1}^{(n+1)} &=&a_{n}\phi _{1}+b_{n}\phi _{2} ;\quad
\phi_{2}^{(n+1)}=c_{n}\phi _{2}+d_{n}\phi _{1}.  \notag
\end{eqnarray}
In order to discuss the asymptotic properties of (\ref{e1}), it is necessary
to determine the coefficients $a_{n}, b_{n}, c_{n}$ and $d_{n}$. After some
straightforward calculation, we can obtain the following relations:
\begin{eqnarray}
a_{n}&=&a_{0}a_{n-1}+a_{n-1}^{\prime }+d_{0}b_{n-1}  \notag \\
b_{n}&=&b_{0}a_{n-1}+b_{n-1}^{\prime }+c_{0}b_{n-1}  \notag \\
c_{n}&=&c_{0}c_{n-1}+c_{n-1}^{\prime }+b_{0}d_{n-1}  \label{e22} \\
d_{n}&=&d_{0}c_{n-1}+d_{n-1}^{\prime }+a_{0}d_{n-1}.  \notag
\end{eqnarray}
Our task is now to introduce the asymptotic aspect of the method. For this
purpose, the $n^{th}$ and $(n+1)^{th}$ derivatives of $\phi_{1}$ and $\phi_{2}$
can be written as
\begin{eqnarray}
\phi _{1}^{(n)} &=&a_{n-1}\left( \phi _{1}+\frac{b_{n-1}}{a_{n-1}}\phi
_{2}\right) , \quad \phi _{2}^{(n)}=c_{n-1}\left( \phi _{2}+\frac{d_{n-1}}{%
c_{n-1}}\phi _{1}\right)  \notag \\
\phi _{1}^{(n+1)} &=&a_{n}\left( \phi _{1}+\frac{b_{n}}{a_{n}}\phi
_{2}\right) , \quad \phi _{2}^{(n+1)}=c_{n}\left( \phi _{2}+\frac{d_{n}}{%
c_{n}}\phi _{1}\right) .  \label{e23}
\end{eqnarray}
The coefficients $d_{0}$ and $c_{0}$ include the coupling constants.
Therefore, for sufficiently large $n$ we can suggest the following
asymptotic constraints:
\begin{equation}
\frac{b_{n-1}}{a_{n-1}}=\frac{b_{n}}{a_{n}}=\lambda _{1}; \quad \frac{d_{n-1}%
}{c_{n-1}}=\frac{d_{n}}{c_{n}}=\lambda _{2}.  \label{e24}
\end{equation}
In this formalism, the relations given in (\ref{e24}) imply that the wave
functions $\phi_{1}$ and $\phi_{2}$ are truncated for sufficiently large $n$
and the roots of the relations given in (\ref{e24}) belong to the spectrum
of the matrix Hamiltonian. Therefore, one can easily compute the energy of
the Hamiltonian by solving (\ref{e24}) for the energy term when $%
z\rightarrow z_{0}$. Under the asymptotic condition of (\ref{e24}), one can
find the wave functions $\phi _{1}$ and $\phi _{2}$. When we take $\frac{%
\phi _{1}^{(n+1)}}{\phi _{1}^{(n)}}$ and $\frac{\phi _{2}^{(n+1)}}{\phi
_{2}^{(n)}}$ by using (\ref{e23}) under the constraints given in (\ref{e24}%
), we obtain:
\begin{equation}
\phi _{1}^{(n)}=\exp \left( \int \frac{a_{n}}{a_{n-1}}dz\right) \quad or
\quad \phi _{2}^{(n)}=\exp \left( \int \frac{c_{n}}{c_{n-1}}dz\right) .
\label{e25}
\end{equation}
Substituting the expression of $a_{n}$ (and $c_{n}$) given in (\ref{e22})
into (\ref{e25}) and then replacing the $\phi_{1}^{(n)}$ (and $\phi_{2}^{(n)}
$) in (\ref{e21}), respectively, one gets the following expressions;
\begin{equation}
\phi _{1}+\lambda_{1}\phi _{2}=\exp \left( \int \left(
a_{0}+\lambda_{1}d_{0}\right) dz\right)\quad or \quad
\phi_{2}+\lambda_{2}\phi _{1}=\exp \left( \int \left(
c_{0}+\lambda_{2}b_{0}\right) dz\right) .  \label{e26}
\end{equation}

Using the second equality in (\ref{e26}), one can substitute $\phi_{2}$
into the first equality in Eq.(\ref{e20}). Thus, one writes
\begin{equation}
\phi _{1}^{\prime}+(b_{0}\lambda_{2}-a_{0})\phi _{1}= C_{1} b_{0} \exp
\left(\int \left( c_{0}+\lambda_{2}b_{0}\right) dz\right)  \label{phi1func1}
\end{equation}
and the solution is found as
\begin{equation}
\phi _{1}=\exp \left(\int \left( a_{0}-\lambda_{2} b_{0}\right) dz\right)
\left[\int C_{1} b_{0}~e^{ \left(\int \left( c_{0}+\lambda_{2} b_{0}\right)
dz\right)} ~ dz + C_{2} \right].  \label{phi1func2}
\end{equation}
If the same procedure is performed for $\phi_{2}$, one finds the solution as
\begin{equation}
\phi _{2}=\exp \left(\int \left( c_{0}-\lambda_{1}d_{0}\right) dz\right)
\left[\int C_{3} d_{0}~e^{ \left(\int \left( a_{0}+\lambda_{1}d_{0}\right)
dz\right)} ~ dz + C_{4} \right].  \label{phi2func1}
\end{equation}

An immediate practical consequence of these results is that the eigenvalues
and eigenfunctions of the various quantum optical Hamiltonians can be
determined. In the following sections, it is shown that this asymptotic
approach opens the way to the treatment of a large class of matrix
Hamiltonians of practical interest.

\section{Results and Discussions}

In this part, we apply the results of previous sections to obtain the
solutions of the Hamiltonians given by Eqs. (\ref{e15a},b) and (\ref{e16a}%
,b). We briefly discuss the corresponding physical system of each
Hamiltonian.

\subsection{Solution of the Hamiltonian $H$ under the constraints: $\protect%
\gamma _{1}=\protect\gamma _{4}=\protect\kappa _{2}=\protect\kappa _{3}=0$}

In this case the Hamiltonian includes two important physical Hamiltonians: $%
E\otimes \varepsilon $ Jahn-Teller (JT) Hamiltonian \cite{reik} and
Hamiltonians of quantum dots including spin-orbit coupling \cite{rashba,koc3}%
. When $\omega _{1}=\omega _{2}=\omega $, $\kappa _{2}=\kappa _{3}=\gamma
_{1}=\gamma _{4}=0$ and $\kappa _{1}=\kappa_{4}=\gamma _{2}=\gamma
_{3}=\kappa$, the Hamiltonian $H$ reduces to the $E\otimes \varepsilon$ JT
Hamiltonian as we have mentioned before. It is obvious that the
corresponding first order differential equations (\ref{e15a},b) for JT
Hamiltonian can be written in the following form:
\begin{equation}
\phi _{1}^{\prime }=a_{0}\phi _{1}+b_{0}\phi _{2};\quad \phi _{2}^{\prime
}=c_{0}\phi _{2}+d_{0}\phi _{1}.  \label{e27}
\end{equation}
where the coefficients $a_{0}, b_{0}, c_{0}$ and $d_{0}$ are given by
\begin{eqnarray}
a_{0} &=&\frac{\kappa ^{2}-2\omega _{0}-2k+2E-2}{4z-\kappa ^{2}}  \notag \\
b_{0} &=&\frac{\kappa \left( \omega _{0}+k+2z+E\right) }{\kappa ^{2}-4z}
\notag \\
c_{0} &=&\frac{\kappa ^{2}(1+k+z)+2z\left( \omega _{0}-k+E-2\right) }{%
z\left( 4z-\kappa ^{2}\right) }  \label{e28} \\
d_{0} &=&\frac{\kappa \left( E-\omega _{0}-k+2z+1\right) }{z\left( \kappa
^{2}-4z\right) }.  \notag
\end{eqnarray}

Using a simple MATHEMATICA program one can compute $a_{n}, b_{n},
c_{n}$ and $d_{n}$ by using the relations given in (\ref{e22}). On
the other hand, for each iteration the expression $
\delta_{1}(z)=b_{n-1}(z)a_{n}(z)-a_{n-1}(z)b_{n}(z)$ (and
$\delta_{2}(z)=d_{n-1}(z)c_{n}(z)-c_{n-1}(z)d_{n}(z)$) depends on
different variables, such as $E_{n},\kappa ,\omega _{0}$ and $z$.
It is also noticed that the iterations should be terminated by
imposing the quantization condition $\delta_{i}(z)=0, i=1,2$ as an
approximation to (\ref{e24}) to obtain the eigenenergies. The
calculated eigenenergies $E_{n}$ by means of this condition
should, however, be independent of the choice of $z$. The choice
of $z$ is observed to be critical only to the speed of
convergence of the eigenenergies, as well as for the stability of
the process. In our study it has been observed that the optimal
choice for $z$ is $z=0$ \cite{amore}. Therefore, we set $z=0$ at
the end of the iterations. We also note that the first value of
the solution set of $\delta_{1}(z)=0$ (or $\delta_{2}(z)=0$) is
not physically acceptable unless the system is exactly solvable.

To fix the iteration number for convergence, the first twenty energy
levels have been determined for $n=8,9,10,11,12,14,16,18$ iterations for the
Hamiltonian above. It has been obtained that $E_{20}=21.103745$ for $n=10$, $%
E_{20}=21.007171$ for $n=11$, $E_{20}=21.007064$ for $n=12$, $%
E_{20}=21.007064$ for $n=13$, and $E_{20}=21.007064$ for $n=14,16,18$
iterations, respectively, for $k=1, w_{0}=0, w=1, \kappa=\frac{1}{10}$. It
is obviously seen that there is no change in the $20^{th}$ energy value for $%
n\geq 12$. Since it is also the same for the other Hamiltonians
given below, $n=14$ iteration is assumed to be sufficient for the
determination of the energy eigenvalues of the related
Hamiltonians.

In Table \ref{JTHam} we have compared our results with previous
studies. It is seen that the results obtained by the AIM agree
with those in Refs.
\cite{longuet,LoPRA5127,kulakCzechJP889,kulakSSC607}.

\begin{table}[t]
\caption{The ground-state energies obtained from Refs.
\cite{longuet,LoPRA5127,kulakCzechJP889,kulakSSC607} and from the
quantization condition by using coefficients given in Eq.
(\ref{e28}) for different values of $\kappa$ in the case that
$k=0, w_{0}=0, w=1$.} \label{JTHam}
\begin{tabular}{cccccc}
\hline\noalign{\smallskip}
$\kappa $ & Ref. \cite{longuet} & Ref. \cite{LoPRA5127} & Ref. \cite{kulakCzechJP889}
& Ref. \cite{kulakSSC607} & Present Results \\
\noalign{\smallskip}\hline\noalign{\smallskip}
0.25 & 0.774 & 0.7766 & 0.7765 & 0.7739 & 0.7738 \\
\noalign{\smallskip}\hline\noalign{\smallskip}
0.5 & 0.578 & 0.5877 & 0.5870 & 0.5780 & 0.5780 \\
\noalign{\smallskip}\hline\noalign{\smallskip}
0.75 & 0.400 & 0.4173 & 0.4158 & 0.3998 & 0.3997 \\
\noalign{\smallskip}\hline\noalign{\smallskip}
1 & 0.233 & 0.2586 & 0.2560 & 0.2331 & 0.2330 \\
\noalign{\smallskip}\hline\noalign{\smallskip}
2 & -0.369 & -0.3157 & -0.3232 & -0.3686 & -0.3689 \\
\noalign{\smallskip}\hline\noalign{\smallskip}
3 & -0.919 & -0.8466 & -0.8575 & -0.9177 & -0.9189 \\
\noalign{\smallskip}\hline\noalign{\smallskip}
5 & -1.961 & -1.8716 & -1.8831 & -1.9540 & -1.9610 \\
\noalign{\smallskip}\hline\noalign{\smallskip}
7 & -2.976 & -2.8833 & -2.8932 & -2.9586 & -2.9760 \\
\noalign{\smallskip}\hline\noalign{\smallskip}
10 & -4.485 & -4.3937 & -4.4019 & -4.4594 & -4.4850 \\
\noalign{\smallskip}\hline\noalign{\smallskip}
15 & -6.991 & -6.9042 & -6.9108 & -6.9610 & -6.9901 \\
\noalign{\smallskip}\hline\noalign{\smallskip}
20 & -9.493 & -9.4111 & -9.4168 & -9.4627 & -9.4809 \\
\noalign{\smallskip}\hline\noalign{\smallskip}
30 & -14.496 & -14.4202 & -14.4249 & -14.4651 & -14.488 \\
\noalign{\smallskip}\hline
\end{tabular}%
\end{table}

\begin{figure*}[h]
\begin{center}
\epsfbox{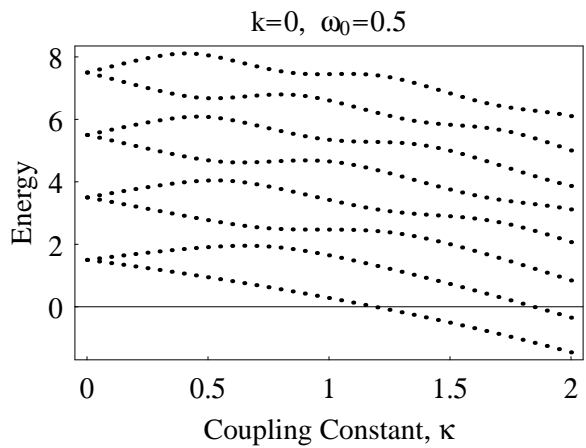}
\end{center}
\par
%\vspace*{5cm}       % Give the correct figure height in cm
\caption{Energy of the displaced coupled harmonic oscillator as a function
of coupling constant.}
\end{figure*}

Under the given conditions the Hamiltonian (\ref{e1}) takes the form:
\begin{equation}
H=H_{0}+\omega _{0}\sigma _{0}+\kappa \left[ \left( a+b^{+}\right) \sigma
_{+}+\left( a^{+}+b\right) \sigma _{-}\right] .  \label{ex27}
\end{equation}%
As a related topic, we mention here that studies of the $E\otimes
\varepsilon $ JT problem led Judd \cite{judd} to discover a class of exact
isolated solutions of the model. To determine the relations between the
parameters of the model, one can obtain an analytic form of two eigenvectors
of the Hamiltonian corresponding to the specific energy. The complete
description of these solutions has been given by Ko\c{c} et al. \cite{koc3}.
They observed that quasi-exact solutions can be obtained by using $osp(2,2)$
super algebra.
\begin{figure*}[h]
\begin{center}
\epsfbox{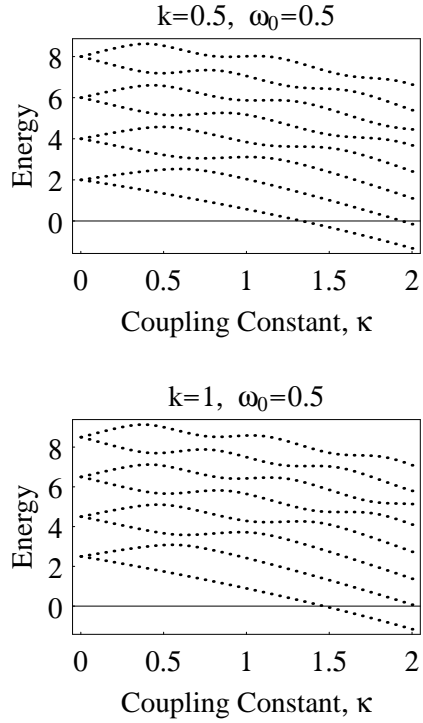}
\end{center}
\par
%\vspace*{5cm}       % Give the correct figure height in cm
\caption{Energy of the octahedral JT systems as a function of coupling
constant.}
\end{figure*}

The physical systems described by the Hamiltonian (\ref{ex27}) are
summarized as follows. When $\omega _{0}=1/2$ and $k=0$, the corresponding
equation related to the Hamiltonian of the displaced coupled harmonic
oscillator whose eigenvalues are obtained as a function of coupling
constant, $\kappa$, is solved by using AIM and the result is given in Figure
1.

The Hamiltonian corresponds to three octahedral JT systems when $\omega
_{0}=1/2$ and $k$ takes integer or half-integer values. These octahedral
systems are $\Gamma _{8}\otimes \left( \varepsilon +\tau _{2}\right) $
linear $E\otimes \varepsilon $ and linear $\Gamma _{8}\otimes \tau _{2}$,
for which their eigenvalues are depicted in Figure 2.

Finally, in the presence of an external field, $\omega _{0}\neq 1/2$, the
Hamiltonian is compatible with the generalized $E\otimes \varepsilon$ JT
system and dimers. For dimers $k=0$ and for generalized $E\otimes
\varepsilon $ JT system, $k$ takes half-integer values. The results of the
iteration for dimers are given in Figure 3 and for generalized $E\otimes
\varepsilon $ JT system are given in Figure 4.
\begin{figure*}[h]
\begin{center}
\epsfbox{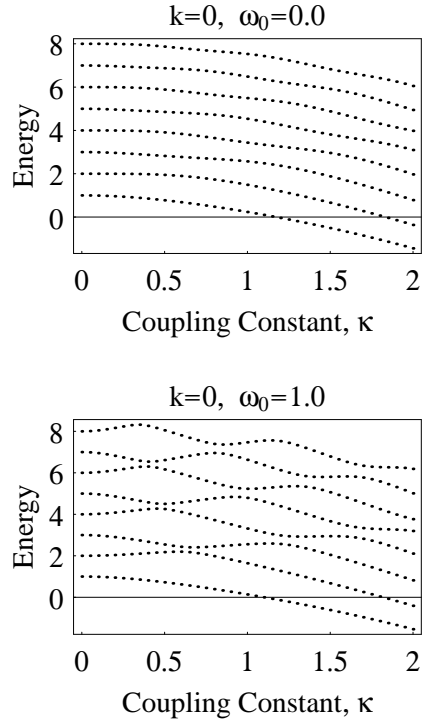}
\end{center}
\par
%\vspace*{5cm}       % Give the correct figure height in cm
\caption{Energy of the dimer as a function of coupling constant.}
\end{figure*}
\begin{figure*}[h]
\begin{center}
\epsfbox{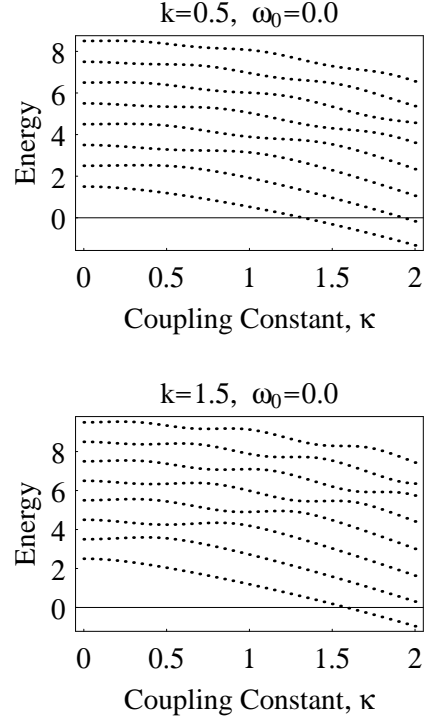}
\end{center}
\par
%\vspace*{5cm}       % Give the correct figure height in cm
\caption{Energy of the generalized $E\otimes \protect\varepsilon $ JT system
as a function of coupling constant.}
\end{figure*}

It is shown in the first 4 Figures that the energy becomes an oscillating
function of the coupling constant $\kappa$ when $k>0$.

The Hamiltonians of quantum dots including Rashba coupling can be obtained
when we set $\omega _{1}=\omega _{2}=\omega $, $\kappa _{2}=\kappa
_{3}=\gamma _{1}=\gamma _{4}=0$ and $\kappa _{1}=-\kappa _{4}=\gamma
_{2}=-\gamma _{3}=\kappa $, and the coefficients of the coupled differential
equations are given by
\begin{eqnarray}
a_{0} &=&\frac{\kappa ^{2}-2\omega _{0}-2k+2E-2}{4z+\kappa ^{2}}  \notag \\
b_{0} &=&\frac{\kappa \left( 2z-\omega _{0}-k-E\right) }{\kappa ^{2}+4z}
\notag \\
c_{0} &=&\frac{\kappa ^{2}+2\left( \omega _{0}+k+E\right) }{4z+\kappa ^{2}}-%
\frac{k+1}{z}  \label{ex28} \\
d_{0} &=&\frac{\kappa \left( E-\omega _{0}-k-2z-1\right) }{z\left( \kappa
^{2}+4z\right) }.  \notag
\end{eqnarray}%
\begin{figure*}[h]
\begin{center}
\epsfbox{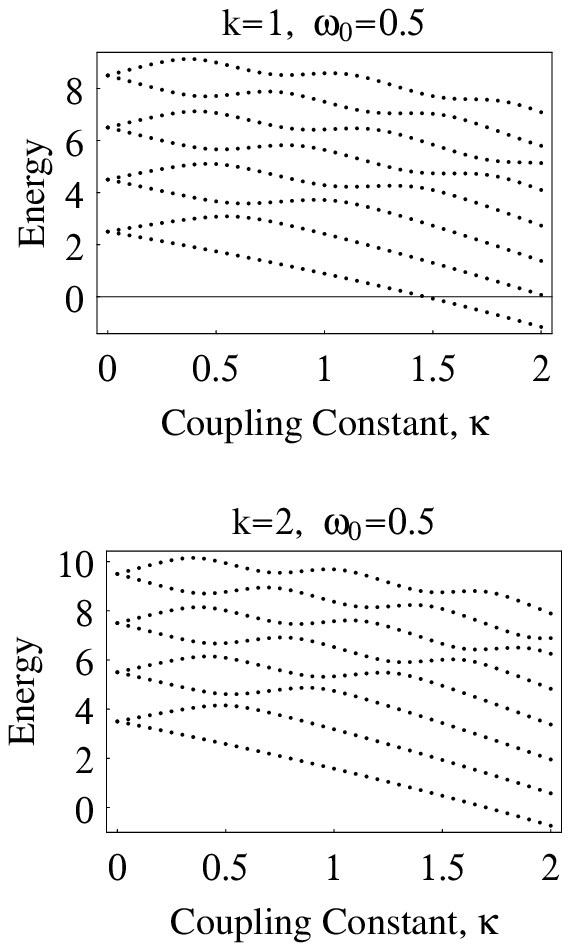}
\end{center}
\par
%\vspace*{5cm}       % Give the correct figure height in cm
\caption{Energy of the Rashba Hamiltonian as a function of coupling
constant. }
\end{figure*}

Note that the origin of the Rashba spin-orbit coupling in quantum dots is
due to the lack of inversion symmetry which causes a local electric field
perpendicular to the plane of the heterostructure. In the literature, the
Hamiltonian has been formalized in the coordinate-momentum space leading to
a matrix differential equation.

\subsection{Solution of the Hamiltonian $H$ under the constraints: $\protect%
\gamma _{1}=\protect\gamma _{3}=\protect\kappa _{2}=\protect\kappa _{4}=0$}

Under the constraints given in this section, we study various well-known
exactly solvable Hamiltonians which give us opportunity to test our
approach. By the given constraint, the Hamiltonian (\ref{e1}) takes the form
\begin{equation}
H=H_{0}+\omega _{0}\sigma _{0}+\left( \kappa _{1}a+\kappa _{3}b\right)
\sigma _{+}+\left( \gamma _{2}a^{+}+\gamma _{4}b^{+}\right) \sigma _{-}.
\label{ex29}
\end{equation}%
We interpret below the solution of the three physical systems using AIM.

\subsubsection{Jaynes Cummings Hamiltonian($\protect\gamma _{4}=\protect%
\kappa _{3}=\protect\omega _{2}=0,\protect\kappa _{1}=\protect\gamma _{2}=
\protect\kappa $)}

The Jaynes-Cummings (JC) Hamiltonian with rotating wave approximation is
given by
\begin{equation}
H=\omega a^{+}a+\omega _{0}\sigma _{0}+\kappa \left( \sigma _{+}a+\sigma
_{-}a^{+}\right)  \label{e29}
\end{equation}
In this case the coefficients of the coupled differential equations (\ref%
{e16a},b) are given by
\begin{eqnarray}
a_{0} &=&-\left( \frac{\kappa^2 + \omega \left( E - k \omega + \omega _{0}
\right) }{\omega^2\,z} \right)  \notag \\
b_{0} &=&\frac{\kappa \left( E - \omega _{0} \right) }{\omega ^2 z}  \notag
\\
c_{0} &=&\frac{-E + \omega + k \omega + \omega _{0}} {\omega z}  \label{e30}
\\
d_{0} &=&\frac{\kappa}{\omega z}  \notag
\end{eqnarray}%
The AIM leads to the following expressions for the eigenvalues of the JC
Hamiltonian
\begin{eqnarray}
n &=&1;\quad E=\left(k + \frac{1}{2} \right) \omega \mp \frac{1}{2} \sqrt{4
\kappa^2 \left(k + 1 \right) + {\left( \omega + 2 \omega _{0} \right) }^2}
\notag \\
n &=&2;\quad E=\left(k - \frac{1}{2} \right) \omega \mp \frac{1}{2} \sqrt{4
\kappa^2 \left(k \right) + {\left( \omega + 2 \omega _{0} \right) }^2}
\notag \\
n &=&3;\quad E=\left(k- \frac{3}{2} \right) \omega \mp \frac{1}{2} \sqrt{4
\kappa^2 \left(k - 1 \right) + {\left( \omega + 2\omega _{0} \right) }^2}
\label{ex31} \\
n &=&n;\quad E=\left( k + \frac{3}{2} - n \right) \omega \mp \frac{1}{2}
\sqrt{4 \kappa^2 \left( k + 2 - n \right) + \left( \omega + 2\omega _{0}
\right)^2 }  \notag
\end{eqnarray}
It is obvious that when the coupling constant $\kappa $ is zero, then the
result is the eigenvalues of the simple harmonic oscillator.

Substituting the variables $a_{0}, b_{0}, c_{0}$ and $d_{0}$ in Eqs. (\ref%
{phi1func2}) and (\ref{phi2func1}), one finds the eigenfunctions $\phi_{1}(z)
$ and $\phi_{2}(z)$ as
\begin{eqnarray}
n &=&1;\quad \phi _{1}=1 \quad \quad \quad \phi _{2}=1  \notag \\
n &=&2;\quad \phi _{1}=z \quad \quad \quad \phi _{2}=z  \notag \\
n &=&3;\quad \phi _{1}=z^{2} \quad \quad \quad \phi _{2}=z^{2}  \notag \\
n &=&n;\quad \phi _{1}=z^{n-1} \quad \quad \phi _{2}=z^{n-1}.
\label{eigenfunc1}
\end{eqnarray}
Using Eq.(\ref{e13d}), one writes
\begin{equation}
\psi (x,y) = x^{k} (y/x)^{n-1}|\downarrow \rangle + x^{k+1}
(y/x)^{n-1}|\uparrow \rangle  \label{psiJC1}
\end{equation}
and finds
\begin{equation}
\psi (x,y) =C x^{1 + k -n} y^{n - 1} \left( |\downarrow\rangle + x |\uparrow
\rangle \right)  \label{psiJCf}
\end{equation}
where $C$ is the normalization constant, $k$ is the state number and $n$ is
the iteration number. In order to find the original eigenfunction of the
Hamiltonian, one can use Eq.(\ref{e14}) by replacing the operators given
in (\ref{e5}) and (\ref{e8}).

\subsubsection{Dirac Oscillator($\protect\omega _{1}=\protect\omega _{2}=$ $%
\protect\kappa _{3}=\protect\gamma _{4}=0)$}

One can also show that the constrains given in this section include the
Dirac oscillator. In order to show this we express the Dirac oscillator with
boson operators. Consider the $(2+1)$ dimensional Dirac equation for a free
particle of mass $m$ in terms of two component spinors, then $\psi$ can
be written as \cite{ito,mosh,koc5}
\begin{equation}
E\psi =\left( \sum_{i=1}^{2}c\sigma _{i}p_{i}+mc^{2}\sigma _{0}\right) \psi .
\label{e32}
\end{equation}
The momentum operator $p_{i}$ is a differential operator $\mathbf{p}=-i\hbar
(\partial _{x},\partial _{y})$ and the 2D Dirac oscillator can be
constructed by changing the momentum as $\mathbf{p}\rightarrow \mathbf{p}%
-im\omega \sigma _{0}\mathbf{r}$. Then the Dirac equation (\ref{e32}) takes
the form
\begin{equation}
\left( E-mc^{2}\sigma _{0}\right) \psi =c\left[ (p_{x}-ip_{y})-im\omega
^{\prime }(x-iy)\right] \sigma _{+}+c\left[ (p_{x}+ip_{y})-im\omega ^{\prime
}(x+iy)\right] \sigma _{-}.  \label{e33}
\end{equation}
After some straightforward treatment, we obtain the bosonic form of the
Dirac oscillator:
\begin{equation}
\left( E-mc^{2}\sigma _{0}\right) \psi =2ic\sqrt{m\omega ^{\prime }\hbar }%
\left[ a\sigma _{+}+a^{+}\sigma _{-}\right] \psi .  \label{e34}
\end{equation}
An immediate practical consequence of these results is the Lie algebraic
structure of the Hamiltonians that can easily be determined. It can be
obtained by setting the parameters of the JC Hamiltonian to $\omega =0, \kappa =2ic%
\sqrt{m\omega ^{\prime }\hbar }$ and $\omega _{0}=mc^{2}$. Then the
eigenvalues of the Dirac oscillator are given by
\begin{equation}
E=\pm \frac{1}{2}\sqrt{4m^{2}c^{4}-4\hbar \omega ^{\prime }mc^{2}\left( k\pm
n\right) }.  \label{ex35}
\end{equation}

\subsubsection{Modified Jaynes-Cummings Hamiltonian$\left( \protect\kappa %
_{1}=\protect\gamma _{2}=\protect\kappa _{3}=\protect\gamma _{4}=\protect%
\kappa ,\protect\omega _{2}=\protect\omega _{1}=\protect\omega \right) $}

In addition to these Hamiltonians, we can also show that our formalism
includes another important Hamiltonian: When a single two-level atom is placed
in the common domain of two cavities interacting with two quantized modes,
the Hamiltonian of such a system can be obtained from the modification of
the JC Hamiltonian and is given by
\begin{equation}
H=\omega a^{+}a+\omega b^{+}b+\omega _{0}\sigma _{0}+\kappa \left(
a+b\right) \sigma _{+}+\kappa \left( a^{+}+b^{+}\right) \sigma _{-}.
\label{e37}
\end{equation}%
Without detailed calculations, one can obtain the energy of the MJC
Hamiltonian in the closed form by using AIM:%
\begin{equation}
E=\left( k+\frac{3}{2}\right) \pm \frac{1}{2}\sqrt{8\left( k+1-n\right)
\kappa ^{2}+\left( 2\omega _{0}-1\right) ^{2}}.  \label{e38}
\end{equation}%
Consequently the exact eigenvalues can be reproduced by AIM.

\section{Conclusion}

We have systematically discussed the solutions of various physical
Hamiltonians within the framework of AIM. We have shown that the formalism
given in this paper leads to the exact or approximate solution of the
problems of various physical systems. We have applied the AIM to the
problem of an electron in a quantum dot in the presence of both a magnetic
field and spin-orbit coupling. The procedure presented here gives an
accurate result for the eigenvalues of the both JT and Rashba Hamiltonians.
The suggested approach can easily be used to solve other quantum
optical problems which are not discussed here.

We have presented a transformation procedure that offers several
advantages, especially if one wishes to describe the eigenvalues of the
bosonic Hamiltonians by using AIM. We have also presented the steps towards
an extension of the AIM.

The technique given in this article can be extended in several ways. The
Hamiltonian of a quantum dot including position dependent effective mass may
be formulated and solved within the procedure given here. We hope that our
method leads to interesting results on the spin-orbit effects in quantum
dots in future studies. Along this line we have work in progress.

\section{Acknowledgements}

The authors would like to thank the referees for their remarkable
suggestions which improved the presentation of the paper. One of the authors
(O. \"{O}zer) is also grateful to the Abdus Salam International Centre for
Theoretical Physics, Trieste, Italy, for its hospitality.

\end{document}